\documentclass[apj, iop]{emulateapj}

\newcommand{\lya}{Ly $\alpha$}

\newcommand{\Hbeta}{\ifmmode {\rm H}\beta \else H$\beta$\fi}
\newcommand{\aliii}{Al\,{\sc iii}}
\newcommand{\oiii}{[O\,{\sc iii}]}

\newcommand{\ovi}{O\,{\sc vi}}
\newcommand{\niv}{N\,{\sc iv}}
\newcommand{\nv}{N\,{\sc v}}
\newcommand{\civ}{C\,{\sc iv}}
\newcommand{\siiv}{Si\,{\sc iv}}
\newcommand{\hei}{He\,{\sc i*}}

\newcommand{\mgii}{Mg\,{\sc ii}}
\newcommand{\feii}{Fe\,{\sc ii}}
\newcommand{\feiii}{Fe\,{\sc iii}}

\newcommand{\Msigma}{\ifmmode M_{\rm BH} - \sigma \else $M_{\rm BH} - \sigma$\fi}
\newcommand{\Mbh}{\ifmmode M_{\rm BH} \else $M_{\rm BH}$\fi}
\newcommand{\mbh}{\ifmmode M_{\rm BH} \else $M_{\rm BH}$\fi}

\newcommand{\Msol}{\ifmmode M_{\odot} \else $M_{\odot}$\fi}
\newcommand{\ergs}{$\rm erg~s^{-1}$}
\newcommand{\kms}{$\rm km~s^{-1}$}
\usepackage{ulem}

\slugcomment{Draft version.}

\shorttitle{ An extreme LoBAL at $z \sim 4.82$}
\shortauthors{Yi et al.}

\begin{document}


\title{The physical constraints on a new LoBAL QSO at $z=4.82$}


\author{
  Weimin Yi$^{1,11,15}$,
  Richard Green$^{2}$,
  Jin-Ming Bai$^{1,11}$,
  Tinggui Wang$^{3}$,
  Catherine J. Grier$^{4}$,
  Jonathan R. Trump$^{4,5,6}$,
  William N. Brandt$^{4,14}$,
  Wenwen Zuo$^{7}$,
  Jinyi Yang$^{8}$,
  Feige Wang$^{8}$,
  Chenwei Yang$^{3}$,
  Xue-Bing Wu$^{8,9}$,
  Hongyan Zhou$^{10}$,
  Xiaohui Fan$^{2}$,
  Linhua Jiang$^{8,9}$,
  Qian Yang$^{8}$,
  Watson Varricatt$^{12}$,
  Tom Kerr$^{12}$,
  Peter Milne$^{2}$,
  Sam Benigni$^{12}$,
  Jian-Guo Wang$^{1,11}$, 
  Jujia Zhang$^{1,11}$, 
  Fang Wang$^{1,11}$, 
  Chuan-Jun Wang$^{1,11,13}$,
  Yu-Xin Xin$^{1,11,13}$,
  Yu-Feng Fan$^{1,11}$,
  Liang Chang$^{1,11}$,
  Xiliang Zhang$^{1,11}$,
  Bao-Li Lun$^{1,11}$
}


\altaffiltext{1}{Yunnan Observatories, Chinese Academy of Sciences, Kunming 650216, China}
\altaffiltext{2}{Steward Observatory, University of Arizona, Tucson, AZ 85721-0065, USA}
\altaffiltext{3}{CAS Key Laboratory for Research in Galaxies and Cosmology, Department of Astronomy, University of Science and Technology of China, China}
\altaffiltext{4}{Department of Astronomy \& Astrophysics and Institute for Gravitation \& the Cosmos, The Pennsylvania State University, 525 Davey Laboratory, University Park, PA 16802, USA}
\altaffiltext{5}{Hubble Fellow}
\altaffiltext{6}{Department of Physics, University of Connecticut, 2152 Hillside Rd Unit 3046, Storrs, CT 06269, USA}
\altaffiltext{7}{Shanghai Astronomical Observatory, Chinese Academy of Sciences, Shanghai 200030, China}
\altaffiltext{8}{Department of Astronomy, School of Physics, Peking University, Beijing 100871, China}
\altaffiltext{9}{Kavli Institute for Astronomy and Astrophysics, Peking University, Beijing 100871, China}
\altaffiltext{10}{Polar Research Institute of China, 451 Jinqiao Road, Shanghai 200136, China}
\altaffiltext{11}{Center for Astronomical Mega-science, Chinese Academy of Sciences, 20A Datun Road, Chaoyang District, Beijing, 100012, P. R. China}
\altaffiltext{12}{UKIRT Observatory, Hilo, Hawaii 96720, USA}
\altaffiltext{13}{University of the Chinese Academy of Sciences, Beijing 100049, China}
\altaffiltext{14}{Department of Physics, Pennsylvania State University, University Park, PA, 16802, USA}
\altaffiltext{15}{Key Laboratory for the Structure and Evolution of Celestial Objects, Chinese Academy of Sciences, Kunming 650216, China}


\begin{abstract}

Very few low-ionization broad absorption line (LoBAL) QSOs have been found at high redshifts to date. One high-redshift LoBAL QSO, J0122+1216, was recently discovered at the Lijiang 2.4-m Telescope with an initial redshift determination of 4.76. Aiming to investigate its physical properties, we carried out follow-up observations in the optical and near-IR spectroscopy. Near-IR spectra from UKIRT and P200 confirms that it is a LoBAL, with a new redshift determination of $4.82\pm0.01$ based on the \mgii~ emission-line.
The new \mgii~ redshift determination reveals strong blueshifts and asymmetry of the high-ionization emission lines. We estimated a black hole mass of $\sim 2.3\times 10^9 M_\odot$ and Eddington ratio of $\sim 1.0$ according to the empirical \mgii-based single-epoch relation and bolometric correction factor. 
It is possible that strong outflows are the result of an extreme quasar environment driven by the high Eddington ratio. 
A lower limit on the outflowing kinetic power ($>0.9\% L_{Edd}$) was derived from both emission and absorption lines, indicating these outflows
 play a significant role in the feedback process to regulate the growth of its black hole as well as host galaxy evolution. 

\end{abstract}


\keywords{broad absorption line quasar: general --- quasar: individual(J0122+1216)}



\section{Introduction}

Extreme gaseous outflows appear most conspicuously in quasar spectra as broad absorption lines (BALs), which typically have velocity widths larger than 2000 km s$^{-1}$, reaching velocities up to tenths of the speed of light.
Previous studies based on spectral analyses indicated that broad absorption line quasars (BAL QSOs) comprise about $\sim$20\% of the quasar population at low and intermediate redshifts \citep{Hewett03,Knigge08,Reichard03b,Scaringi09},  usually characterized by high-ionization broad absorption lines (HiBALs) (e.g., \civ, \siiv, \nv, \ovi). 
A small fraction ($\sim$10\%) of the BAL QSOs also shows absorption troughs characterized by low-ionization species (\mgii $\lambda$2798, \aliii$\lambda$1857) in their spectra. These are called LoBALs (in contrast to HiBALs, which only show high-ionization absorption troughs). An even rarer family of BALs are FeLoBALs, which usually show prominent \feii\ and \feiii\ absorption. 
In general, high-ionization absorption troughs tend to be broader than the troughs associated with low-ionization species (e.g. \citealt{Filiz14}), suggesting quasar outflows have a wide range of ionization states. Recently, the discovery of ultra-fast outflows ($v > 0.1c$, e.g., \citealp{Rogerson15}) presented the phenomenon that the \civ\ absorption trough overlaps with that of \siiv, and \siiv\ with \nv, thus the outflow redshift is often hard to establish without aid from wide wavelength coverage spectra.

In low-redshift cases, the reddening of normal quasars in general can be well fit by SMC-like extinction curves, while more heavily absorbed AGNs are often best characterized by a flatter extinction curve (e.g., \citealp{Maiolino01, Maiolino04a, Gaskell04, Gallerani10}). 
In contrast, reddening in BAL QSOs at high redshifts may deviate from the SMC extinction curve, perhaps due to dust absorption associated with an extreme evolutionary phase \citep{Gallerani10}.
Some studies found that HiBAL quasars are on average more reddened than non-BAL quasars. LoBAL quasars are more reddened than HiBAL quasars (e.g., \citealp{Sprayberry92}), and FeLoBAL quasars are more reddened than LoBAL quasars (e.g., \citealp{Reichard03b}).
BAL QSOs are usually characterized by redder continua compared with non-BAL QSOs, which is commonly interpreted as reddening by dust associated with the outflowing gas (\citealt{Reichard03b}). 
BAL QSOs are some of the most reddened objects observable at high redshift, and so they are useful for understanding internal extinction in the early Universe (e.g.,  \citealt{Gallerani10} )

Originally, the difference between BAL and non-BAL QSOs was mainly ascribed to orientation effects with otherwise similar intrinsic structure
 \citep{Weymann91, Goodrich95, Hines95, Gallagher07}. 
However, one dedicated study on high-redshift BAL QSOs found systematic differences between BAL and non-BAL QSOs with respect to observational properties that should be isotropic, which indicated that these two classes of quasars may be caused by intrinsically different physics to some extent \citep{Gallerani10}. 
No matter how many interpretations are proposed, the basic understanding of all the BAL phenomena is that some dense material flows out from the center of the active galactic nucleus (AGN), which can be detected and characterized by observations from IR to X-ray wavelengths.  
For instance, the general trends that the radio morphology of BAL QSOs is more compact than those of non-BAL QSOs \citep{Becker00} and that BAL QSOs have UV to soft X-ray flux ratios 10 - 30 times smaller than unabsorbed quasars \citep{Brandt00}, greatly advanced our understanding of inner physics of the two kinds of QSOs. 
Dust-reddened quasars, which are plausibly in an extremely active phase, appear likely to host LoBALs (\citealt{Urrutia09}), indicating the LoBAL phenomenon may be linked to evolution, though the orientation effects cannot be ruled out. Some BAL quasars, especially LoBALs, may be cocooned by dust and gas rather than being quasars with a particular line of sight through their disk winds (\citealt{Becker00}).

Variability studies give additional insight into the nature of BAL QSOs. 
For example, \citet{Filiz13} presented a comprehensive study on the variability of \civ~ and \siiv~ BALs over multi-year epochs in a large quasar sample, in which they suggested that global changes in ionization state are likely to be the most straightforward mechanism for explaining  coordinated variability of multiple \civ~ troughs at different velocities. However, some variability of LoBALs may be caused by a BAL structure moving in/out of our line of sight to the UV continuum emitting region (e.g., \citealp{Hall11,Vivek14}).
Variability of BAL trough strengths is relatively common (e.g., \citealp{Hall07,Gibson08,Filiz13}), though only a few cases of dramatic changes ($>$50 \%) have been reported \citep{Crenshaw00, Lundgren07, Hamann08, Leighly09, Gibson10, Filiz12,Grier15,Rogerson15, Vivek16}. 
Yet it is difficult to find a single trigger for BAL variability, since it may be caused by either gas motion across the line of sight or changes in the ionization state (e.g., \citealp{Vivek14,WangT15}).

In this paper, we investigate the physical properties of a newly discovered high-redshift LoBAL QSO using optical and near-IR spectroscopic observations. 
Such distant and powerful outflows are generally believed to exert significant feedback effects on the formation of host galaxies as well as the regulation of supermassive black holes (SMBHs) (e.g., \citealp{Kauffmann00, Richards02, Di05, Everett05,Reeves09,Rupke11,Fabian12}). 
The study of this newly discovered LoBAL QSO at z $\sim$ 5 has the potential to  yield information on whether or not these outflows are energetic enough to provide feedback to its host galaxy.
Though LoBAL QSOs are not a large population, their extreme properties may provide more stringent and valuable tests of BAL outflow models and feedback processes than do the more numerous \lq\lq{normal}\rq\rq~ BAL QSOs.

This paper is organized as follows. In Section 2, we describe the observations from the Lijiang 2.4m telescope (LJT) and UKIRT and the related data reduction we have performed. Through the analyses of all the spectra collected together, we illustrate the spectral properties of this LoBAL QSO, and estimate its black hole mass, Eddington ratio, extinction and the outflow kinetic power in Section 3. The results and implications are discussed further in Section 4, and summarized in Section 5. A flat cosmology with $H_0$ = 70 km s$^{-1}\cdot$Mpc$^{-1}$,  $\Omega_M$ = 0.3 and  $\Omega_{\Lambda}$ = 0.7 has been adopted unless stated otherwise.

\section{Observations and Data Reduction}

From new optical-IR selection criteria based on SDSS and {\it WISE} photometric data \citep{wxb12,wangf16}, which have been demonstrated to select $z \sim 5$ quasars with both high efficiency and completeness, we discovered a high-redshift BAL QSO, SDSS J012247.34+121624.0 (hereafter J0122+1216, \citet{Yi15}), with Balnicity Index (BI) $\sim$16000 \kms (possibly a lower limit) using the LJT on 2014 October 24. As BAL QSOs with large BI and high luminosity are rare (\citealp{Vestergaard03,Sulentic06,Trump06, Gibson09}), we performed follow-up observations including near-IR spectroscopy with TNG/NICS and UKIRT/UIST, and optical spectroscopic monitoring observations from the Yunnan Faint Object Spectrograph and Camera (YFOSC) mounted on the LJT (see Table \ref{table1}). All the spectra obtained were first corrected for Galactic extinction \citep{Schlafly11} with an average Galactic extinction law ($R_V$ = 3.1) and transformed into the quasar rest frame using the redshift determined by the \mgii\ 
emission-line. 
\begin{table}[h]
\centering
 \caption{Near-IR and optical monitoring observations}
 \begin{tabular}{lcccr}
  \hline\noalign{\smallskip}
Instrument & $\lambda / \Delta \lambda $ &  Slit  &  Integration  &  Observation  \\
Name &  &  Width  &  Time  &  Date \\
           &  &  (arcsec)  &  (seconds)  &  (UT) \\
  \hline\noalign{\smallskip}
TNG/NICS & 500 & 1.0 & 800$\times$2  & 2015-02-13  \\ 
UKIRT/UIST & 350 & 0.8 & 240$\times$20 & 2015-10-30   \\
LJT/YFOSC & 320 & 1.8/2.5 & 2400-2700 & 2014 to 2016   \\
P200/TripleSpec & 2700 & 1.0 & 300$\times$8 & 2017-02-09   \\
  \noalign{\smallskip}\hline
\end{tabular}
\label{table1}
\tablecomments{The first near-IR spectrum was obtained by TNG/NICS, but with poor S/N. Later, we obtained its near-IR spectra from UKIRT/UIST and P200/TripleSpec, respectively. Optical spectrophotometric observations were carried out by LJT/YFOSC. }
\end{table}


\subsection{Optically spectrophotometric observations}

Optical spectrophotometric monitoring observations of this BAL QSO were obtained with LJT/YFOSC. Among its various imaging and spectroscopic observing modes \citep{Fan15}, YFOSC provides high-sensitivity, low-resolution spectroscopy, and a quick switch between imaging and spectroscopic modes when observing this faint quasar. 
Typical FWHM in the YFOSC data vary in the range from 0.8\arcsec to 2.0\arcsec. Each night the standard calibration data, including bias, sky-flat fields, and internal lamp flats were obtained. Exposures of the Neon and Helium lamps were used for wavelength calibration. 

\begin{table}[h]
\centering
 \caption{Magnitudes from multiple surveys}
 \begin{tabular}{lcr}
  \hline\noalign{\smallskip}
 Name & J0122+1216 & Reference star \\
 Ra (J2000) & 01:22:47.34  & 01:22:57.8 \\
 Dec(J2000) & +12:16:24.0 & +12:15:42.0\\
 SDSS-u & 23.35 $\pm$0.62 & 18.88$\pm$0.02 \\
 SDSS-g &  24.23$\pm$0.41 & 17.47$\pm$0.01\\
 SDSS-r  &  22.27$\pm$0.16  & 16.91$\pm$0.01\\
 SDSS-i & 19.36$\pm$0.02  & 16.73$\pm$0.01\\
 SDSS-z & 19.24$\pm$0.06 & 16.64$\pm$0.01\\
 Y & 18.484$\pm$0.036 & 16.039$\pm$0.006\\
 J &  17.979$\pm$0.043 &  15.751$\pm$0.006\\
 H & 17.183$\pm$0.042 &  15.326$\pm$0.008 \\
 K & 16.604$\pm$0.042 &  15.283$\pm$0.012\\
 W1 & 15.58$\pm$0.053 &  15.231$\pm$0.044\\
 W2 & 15.03$\pm$0.095 & 15.321$\pm$0.121\\
 W3 & 11.50$\pm$0.165 &  12.701 \\
 W4 & 8.62 &  9.184\\
  \noalign{\smallskip}\hline
\end{tabular}
\label{table2}
\\
\tablecomments{Some of W3 and W4 have no error bar}
\end{table}

Photometric images of J0122+1216 were obtained before spectroscopic observations to select a nearby non-variable reference star, whose spectra could be obtained simultaneously with those for J0122+1216. We also select reference stars from the Catalina Survey \citep{Drake09} in the same field view of YFOSC. 
As a result, an F/G type star (see Table \ref{table2}) was chosen as the best candidate according to its little photometric variation between 2006 Sep and 2013 July. The uncertainties of the photometric measurements include the fluctuations due to photon statistics and the scatter (of 0.03 mag or less) in the measurement of the constant-flux dwarf used. The typical total uncertainty is in the range of 0.01-0.05 mag depending on object brightness and the observing conditions. 
Therefore, our QSO monitoring campaign can offer an efficient and reliable way to search for possible links between emission-line and BAL-trough variability, which may provide clues into the origin of BAL outflows. 
The photometric magnitudes of J0122+1216 and the reference star obtained from the SDSS, UKIDSS, and {\it WISE} databases are listed in Table \ref{table2}.

\begin{figure}[h]
 \includegraphics[width=8.5cm, height=11cm, angle=0]{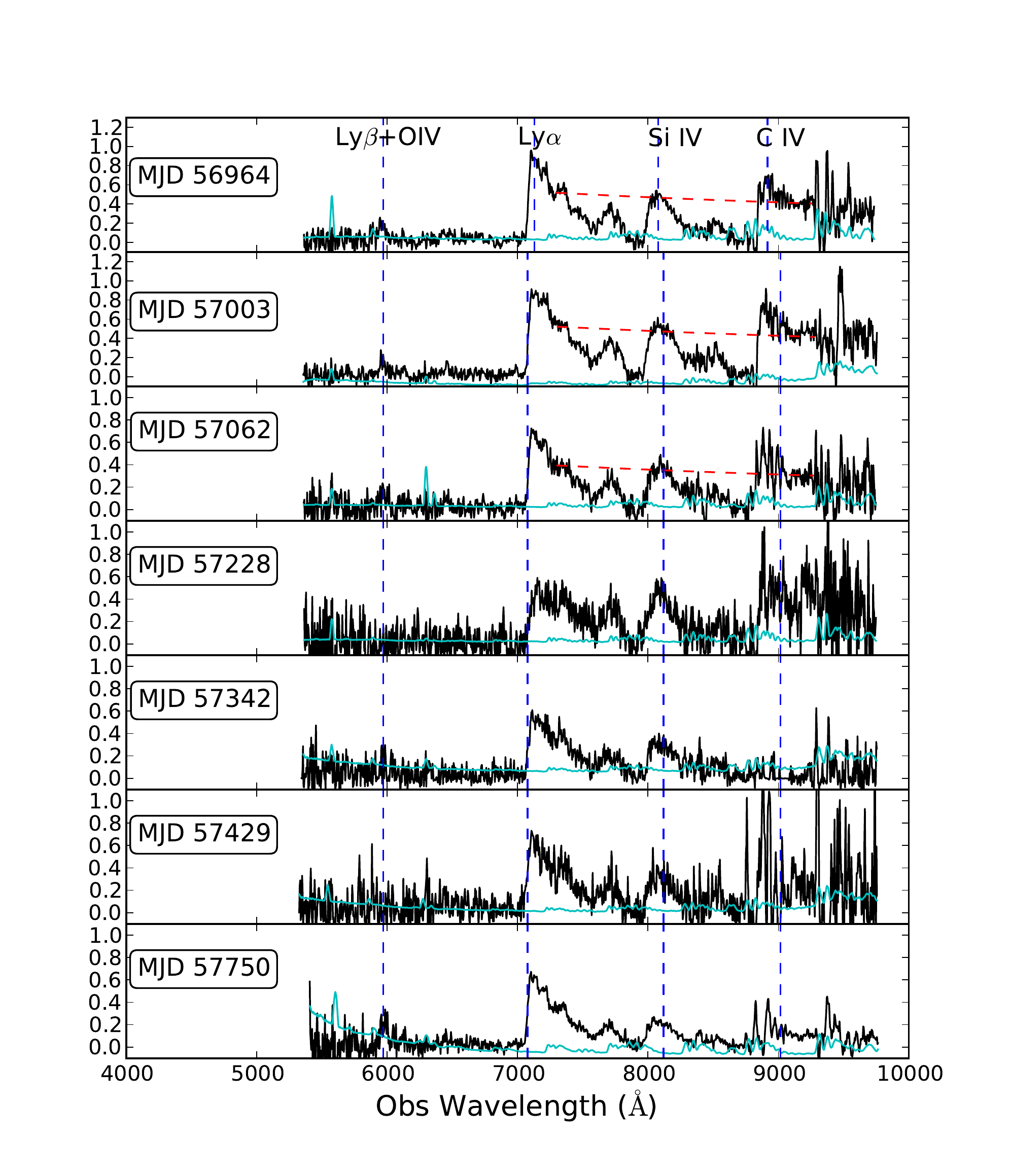}
 \caption{ Optical spectra of the LoBAL obtained by LJT/YFOSC during the past 26 months, for which we did continuum fits ( red dashed lines ) for 3 spectra with relatively high S/N to recover deeply absorbed regions (\civ~ and \siiv) based on a fairly matched comparison spectrum at low-redshift \citep{Zhou10}. Vertical blue dashed lines mark approximate positions for the Ly$\beta$, Ly$\alpha$, \siiv, and \civ~ emission lines. The cyan lines show spectral errors from data reduction.  }
  \label{fig1}
\end{figure}

During the optical spectroscopic observations, we obtained more than one spectrum for each exposure. To achieve good spectrophotometry, the spectrograph's focal plane is rotated to an appropriate position angle so that a nearby reference star could be simultaneously located in the slit. This method allows spectrophotometric calibration of the quasar even under non-photometric conditions. The simultaneous observations reduce the effects of atmospheric dispersion at non-parallactic sky angles, as well as light losses due to occasional guiding errors or poor seeing. 
Exposure times ranged from 2400s to 2700s for each spectroscopic run (see Fig. \ref{fig1}) depending on the transmission and seeing, to aim for a S/N$>$10 pixel$^{-1}$ in the continuum. The spectroscopic data were reduced using standard IRAF routines. The images were bias and flat field corrected. The extraction width is typically 10 pixels, and wavelength calibration is applied to data after the spectral extraction. This procedure results in two spectra from each image: one for J0122+1216 and the other for the comparison. The consecutive quasar/star flux ratios are then compared to correct for systematic errors in each observation and to identify and further remove cosmic-ray events. The ratio is usually reproducible to 0.5\% - 5\% at all wavelengths, 
and observations with ratios differing from the mean value by more than 5\% are discarded. 
With the spectrophotometric runs carried out with LJT/YFOSC, we got 7 optical spectra of the target, as shown in Table \ref{t3}. The spectra from LJT/YFOSC cover the wavelengths from 5500 to 9700 \AA~ at average resolution $R = \lambda/\Delta\lambda \approx$ 320.

\begin{figure}[h]
 \includegraphics[width=8.5cm, height=8cm, angle=0]{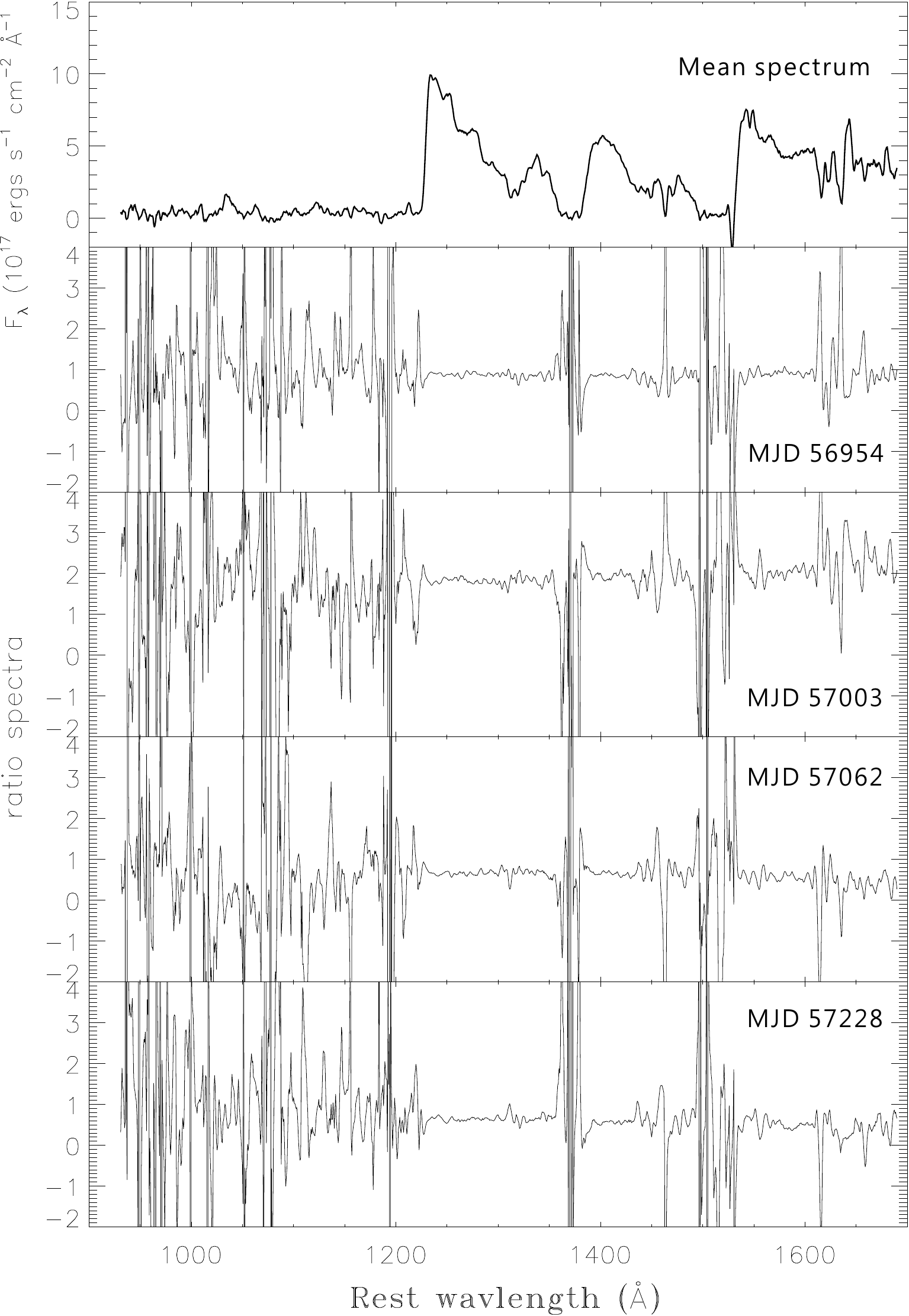}
  \caption{The top panel of this figure shows a mean spectrum of J0122+1216, which is taken as a reference to produce 4 ratio spectra plotted below as a demonstration.}
  \label{new_fig2}
\end{figure}

To determine variability as well as maximum and minimum velocities of the \civ\ and \siiv\ absorption troughs, we smoothed the mean continuum-normalized spectrum using a boxcar smoothing algorithm over three pixels. 
In order to measure the intrinsic variability of the broad absorption troughs reliably rather than the overall variability of this LoBAL quasar, we made different local fits for each individual spectrum according to the neighboring line-free regions of unabsorbed continua when calculating their Balnicity Index (see more in Section 3.2). To be specific, we first take the unabsorbed parts of the \siiv\ and \civ\ emission lines as a reference to recover the continuum in the optical spectra. 
Moreover, we found that the broad band SED of J0122+1216 is similar to that of a quasar (SDSS J100713.68+285348.4) from \citet{Zhou10};  thus the continuum level could be estimated using the latter as a comparison template to recover the absorption-free spectrum. In addition, a mean spectrum of J0122+1216 and ratio spectra compared to the mean composite are obtained as shown in Fig. \ref{new_fig2}. 
At a first glance, the emission/absorption lines appear to vary to some extent during these times. This way of determining variability of emission/absorption lines, however, yields large uncertainties for our lower-S/N spectra and is affected by sky background contamination. 

\begin{figure}[h]
 \includegraphics[width=9cm, height=7cm, angle=0]{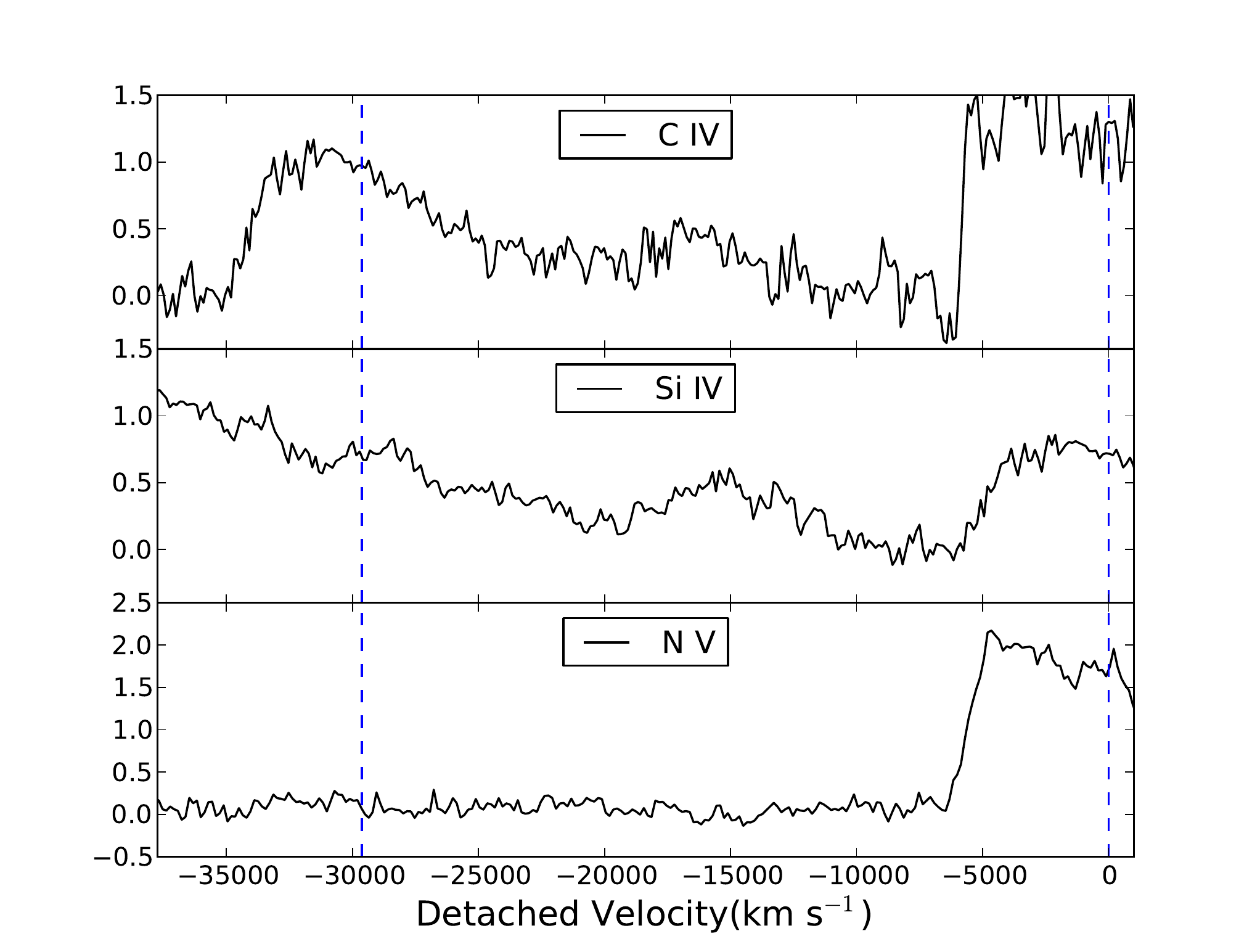}
  \caption{The velocity plots for \civ, \siiv, \niv~ absorption lines. We use the blue dashed lines to mark blueshifted absorption regions for these species. The 29000 \kms~ maximum velocity is set to avoid blending other lines in the absorption region.}
  \label{new_fig1}
\end{figure}
Fortunately, one feasible way to alleviate uncertainties from continuum fitting is to study the flux ratios between our target and the comparison star from simultaneous spectroscopic observations during different epochs. When using the spectra of reference star to recover the continuum level for all the spectra of J0122+1216 obtained from the monitoring campaign, the variability of the \civ\ absorption troughs is  not detected (less than 10\%) , which is consistent with most low-redshift BAL QSOs previously studied. 
Our flux calibration is robust due to simultaneous observations for the non-variable star within the same slit. This also minimizes any possible spectral variability that may be caused either by system fluctuations or weather changes. 

\begin{table}[h]
\centering
 \caption{Optical monitoring observations}
 \begin{tabular}{lccccr}
  \hline\noalign{\smallskip}
Obs & Ly$\alpha$/\siiv~ &  BI(\civ)  &$L^b$&  Exp  &  Ly$\beta$  \\
Date & ratio$^a$ & & $10^{45}$  &Time & EW$^c$  \\
        MJD   &  &  (\kms)  & \ergs &  (s)  &   \\
  \hline\noalign{\smallskip}
56954 & 1.74 & 15760 & 6.3 & 2400  & -15.8  \\ 
57003 & 1.66 & 16090 & 6.7 &2400 & -8.4   \\
57062 & 1.61 & 15050 & 6.1 &2400 & -13.3 \\
57228 & 0.97 & - & - & 2700 & - \\
57342 & 1.42 & - & - &2400 & - \\
57429 & 1.91 & - & - &2100 & -  \\
57750 & 2.50 & - & - &2700 & -26.0  \\
  \noalign{\smallskip}\hline
\end{tabular}
\label{t3}
\tablecomments{Rest-frame UV variabilities based on the spectral analyses. Optical spectrophotometric observations were carried out with LJT/YFOSC  7 times from 2014 Oct to 2016 Dec. \\
a: The ratio between peaks of the two lines\\
b: The \civ\ emission line luminosity \\
c: EWs are derived by neighbor regions around the emission line}
\end{table}

We then determined trough velocities by including only regions with flux below 90\% of the estimated continuum level (i.e., where the normalized flux density is $< 0.9$). A maximum velocity of $\sim 29000$ \kms~ was set to avoid interference by broad lines since outflows at such high velocities are very rare in the UV band. It is clear that both \civ~ and \siiv~ have similar velocity structures, which is particularly  characterized by their two deep absorption troughs. For \nv~, a shallow absorption trough seems to be imprinted at $\sim 2000$ \kms , but it is impossible to obtain other absorption features at higher velocities due to the heavily absorbed \lya~ forest. In addition, the \siiv\ and \civ\  BALs have a similar trough velocity width of 23000 \kms .   
The maximum and minimum velocities of the \siiv\ and \civ\ troughs in each epoch were measured separately, and were consistent over all epochs within the wavelength calibration uncertainty.

\subsection{Near-IR spectroscopic observations}

\begin{figure*}[!]
  \includegraphics[width=19cm, height=5cm]{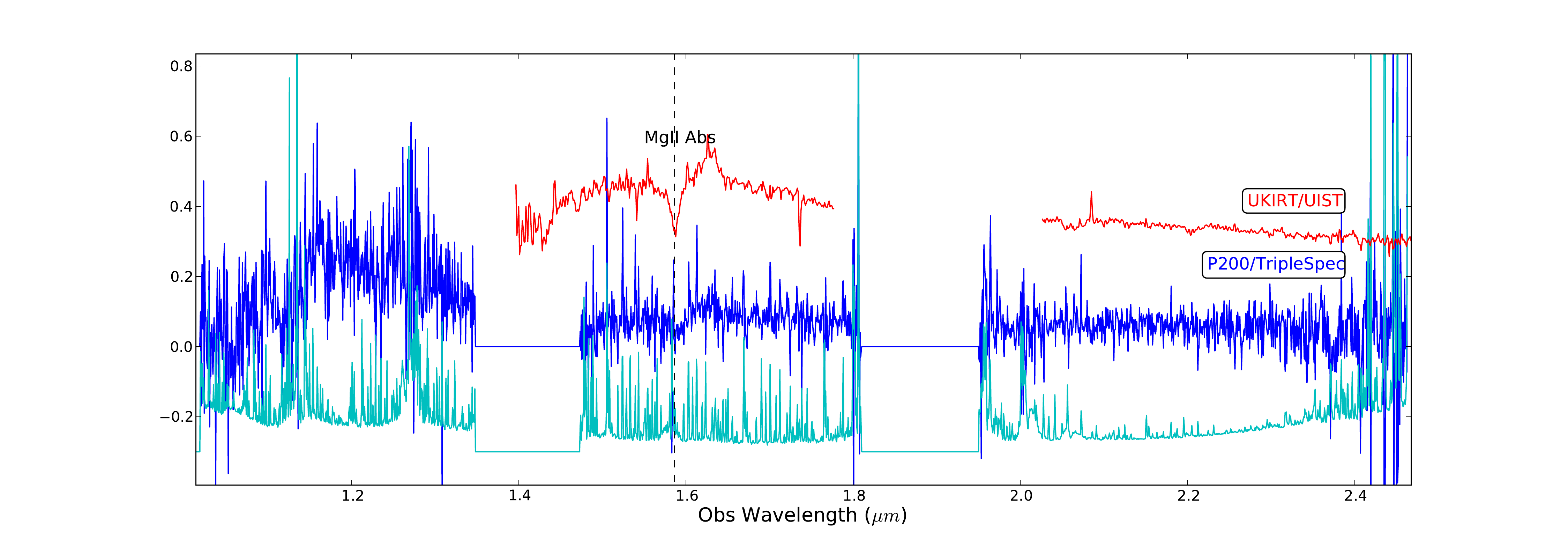}
  \caption{ Near-IR spectra of J0122+1216 were obtained by UKIRT/UIST (red line) and P200/TripleSpec (blue line, smoothed by 5-pixel boxcar filter), respectively. The cyan line is the spectral error distribution from P200/TripleSpec. The vertical black dashed line mark the \mgii~ absorption position.}
  \label{fig2add}
\end{figure*}
With director's discretionary time from the TNG observatory, the first near-IR spectrum was obtained by TNG/NICS on 2015 Feb 13 through service mode. Unfortunately, the S/N of the raw spectrum was too low to be analyzed due to poor seeing and transparency plus large airmass effects. Successful near-IR spectroscopic observations were carried out on 2015 Oct 30 in Hawaii using UKIRT/UIST with an ABBA dither pattern, and each exposure was set to 240 seconds to subtract optimally the sky emission lines. 
Based on the analysis of the performance of UIST and the fact that this target is faint in the JHK bands, we adopted a 0.8\arcsec slit and HK grism, which covers 1.395$\mu$m  to $2.506\mu$m with average spectral resolution of $R \sim$ 350. 
Bias, flat-field and standard-star observations were also taken before and after exposures of the target. 
The IR observations had excellent seeing ($\sim 0.4$ \arcsec~ in H-band) and low humidity ($\tau \sim 0.05$).
The data-reduction was automatically performed using the UIST pipeline based on Starlink packages.

Because the spectral resolution from UKIRT/UIST is low, another high-resolution near-IR spectrum covering the JHK bands was obtained from P200/TripleSpec on Feb 9, 2017 for the purpose of investigating further details. In order to obtain a higher S/N spectrum of J0122+1216, we took 10 Fowler depths to optimize the readout noise after each single exposure using the ABBA mode. The total exposure time was initially assigned to be 5400s, but we then reduced the time to  $300\times8$s when noticing the S/N of each single spectrum started to weaken. This is probably due to the average seeing becoming larger than 1.5\arcsec~ and clouds starting to increase. The data reduction of all spectra was performed using Spextool packages provided by the Palomar Observatory website. The S/N of the combined spectra from P200/TripleSpec was low; however, this spectrum can be used for determining the redshift associated with that from  UKIRT/UIST when noticing they have similar \mgii~ emission/absorption profiles (see Fig. \ref{fig2add}).

\begin{figure}[h]
 \includegraphics[width=9.0cm, height=3cm, angle=0]{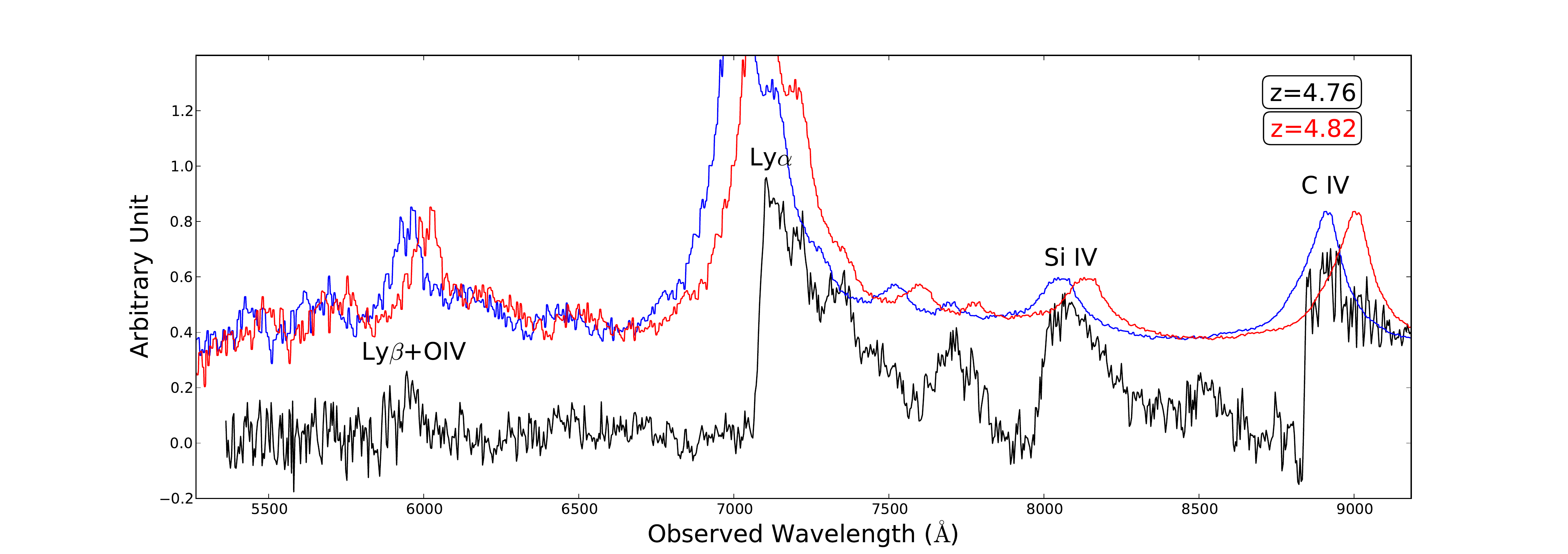}
  \includegraphics[width=9.0cm, height=3cm, angle=0]{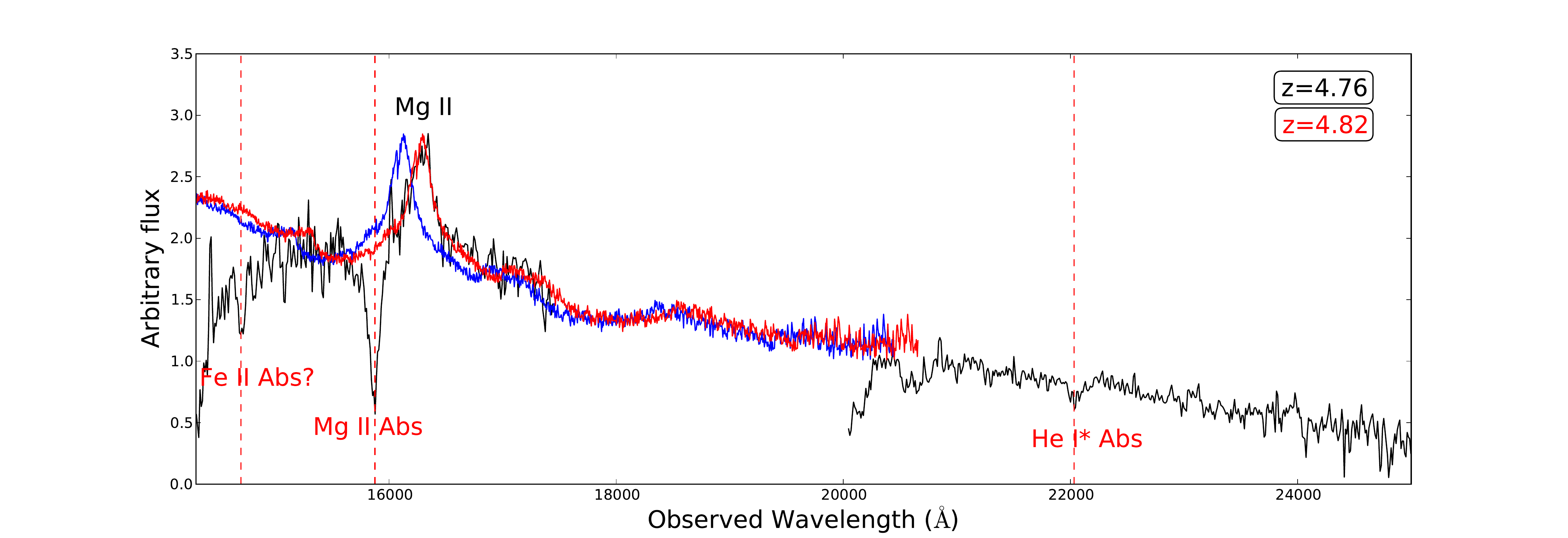}
  \caption{The optical and near-IR spectra (black line) of J0122+1216 obtained by LJT/YFOSC and UKIRT/UIST. For comparison, the composite templates with two different redshifts are shown with blue and red lines, among which we could see the apparent inconsistency. We marked three low-ionization absorption features by red vertical dashed lines.}
  \label{fig2}
\end{figure}

With the aid of the near-IR spectrum obtained by UKIRT/UIST and P200/TripleSpec, this quasar is confirmed as a LoBAL QSO with the largest BI(\civ) $\sim 16000$ \kms~ at $z>3$, matching more tentative observations in the discovery paper. Some low-ionization absorption lines (red dashed lines in Fig. \ref{fig2}) such as \mgii~ and \hei\ were also found with greater than 2$\sigma$ significance, while only tentative evidence for other lines like Fe II is found (i.e., they are found at less than 2$\sigma$ significance). These low-ionization absorption lines, in turn, provide additional diagnostics to study outflow properties.




\section{Spectral properties and analyses}
Through a comprehensive analysis of \civ\ $\lambda$1549, previous studies revealed that blueshifted emission lines associated with outflows in radio quiet quasars are more common and larger amplitude at high luminosity (e.g., \citealp{Wang11,Marziani16}). 
Due to the blueshifts of \civ~ and other high-ionization emission lines, the low-ionization \mgii~ emission-line is more useful for estimating the systemic redshift
 \citep{Richards02,Hewett10,Shen16}. 
This argument has been further supported by an observed symmetry of the \mgii\ emission-line. 
The \mgii\ line in our near-IR spectrum is consistent with being symmetric, leading us to conclude that it is a reliable indicator of the systemic redshift of J0122+1216.

The initial redshift of $z = 4.76$ was determined by \citealt{Yi15} based on optical spectra using the Ly$\alpha$, Ly$\beta$, \civ\ and \siiv\ emission lines matched to the average composite template, while our new redshift determined from the \mgii\ emission-line in the near-IR spectroscopy is 4.82$\pm 0.01$. With the latter redshift, the unabsorbed emission line peaks of Ly$\beta$, \civ\ and \siiv\ exhibit blueward shifts, which are commonly seen among low-redshift BAL QSOs (e.g., \citealp{Richards02,Wang11}), but most of them are rarely with such a dramatic blueshift. This scenario, in fact, can be further corroborated with the evidence of significant variability among these emission lines (see Table \ref{t3}), which is discussed more in the following sections. 

Up to now, there is no widely accepted definition to classify LoBAL or FeLoBAL QSOs with respect to the width and depth of the \mgii\ or \feii\ absorption lines 
(e.g., \citealp{Trump06, Gibson09, Zhang10}). At wavelengths shortward of 14500 \AA, the increasing atmospheric opacity makes detailed line identifications difficult. 
With this caveat in mind, the near-IR spectrum has potential \feii\ absorption features (e.g., the UV62 and UV63 components in the Fe absorption atlas and shallow \feii\ UV1 absorption at $\sim$ 2600 \AA~; see \citealt{Hall02}). Alternatively, it may be experiencing a transition phase between FeLoBAL and LoBAL QSOs characterized by the emergence or disappearance of \feii/ \feiii\ absorption lines. 
The possibility of blended Fe emission/absorption lines makes the measurement of its continuum more uncertain in this spectral region. 
\begin{figure}[h]
 \includegraphics[width=9cm, height=6.5cm, angle=0]{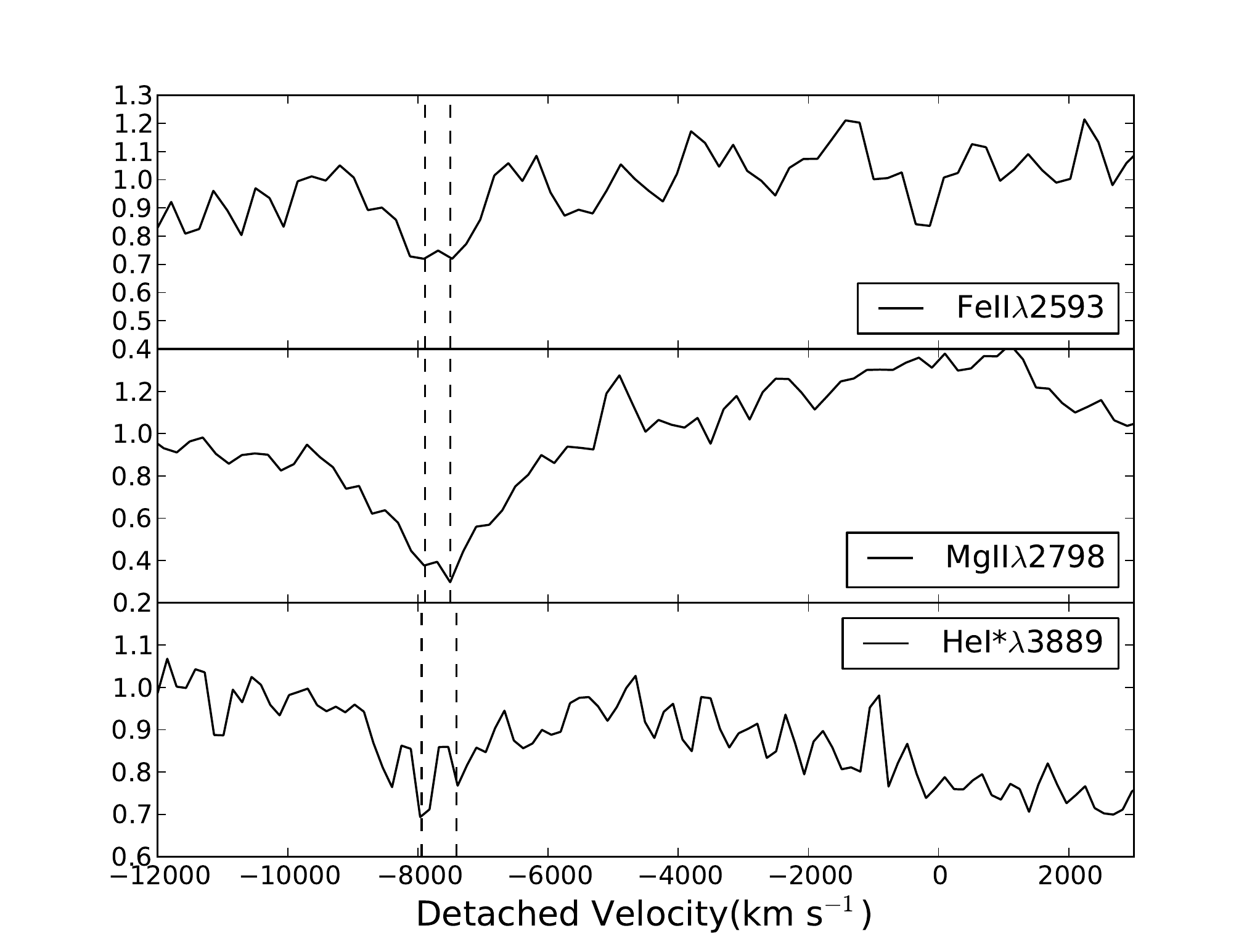}
 \caption{Spectral regions associated with \feii~, \mgii~, \hei~ at the same common blueshift velocity. We use the negative sign to mark the detached velocity on the blueward, which is derived from the systemic redshift of 4.82. The vertical dashed lines show similar velocity components, and y-axis is an arbitrary unit. }
  \label{fig:simple3}
\end{figure}

Adopting the method from \citet{LiuW15}, we quantitatively analyzed the reliability of different absorption features in the near-IR spectrum. 
All three of the low-ionization absorption lines have velocity widths larger than 1600 \kms. Among them, the \mgii\ and \hei\ absorption lines are detected at 3.2$\sigma$ and 2.5$\sigma$ significance, respectively, 
while \feii\ was only detected at 0.9$\sigma$ significance and should thus be treated with caution. However, we argue that the \feii~ absorption feature is likely to be real, given that the \feii\ absorption trough peaks lie at the same position in the common velocity absorption system with respect to the \mgii~,  \hei\ and \feii\ outflowing ions (Fig. \ref{fig:simple3}).
These absorption lines have similar detached velocities ($\sim 8000$ \kms) and velocity widths ( $\sim 2000$ \kms). Such observational properties would be useful for setting physical constraints with respect to the origin of outflows. However, a more detailed study of these ions would require spectra of higher S/N and resolution.

To determine whether this newly discovered LoBAL QSO is typical when compared to the general BAL QSO population, we compare its properties with those of the BAL QSO sample of \citet{Gibson09}. J0122+1216 has obvious emission/absorption blueshifts that are commonly seen among the BAL population, but J0122+1216 itself is less reddened than typical BAL QSOs. Moreover, it is clear to see the steep onset of the \civ\ and Ly$\alpha$ absorptions, which may be caused by inherently similar geormetry and physics of absorption. The velocity width of the \civ\ trough is not unusual, but is significantly larger than 90\% of \civ\ BAL troughs \citep{Gibson09}. Though the velocity of the \civ\ trough falls within the normal range of the BAL QSO population, the maximum velocity of the trough falls toward the high end of the distribution. In general, neither the emission nor its absorption features are strong outliers in the observational properties examined, though they do fall in less common regions with respect to the BI index, the velocity width and range.

\subsection{Estimated BH mass based on the \mgii\ line}
The single-epoch \mgii~ estimator derived from reverberation mapping has been applied to the study of large samples of AGN at $1.5 \le z \le 5$, and it is also preferentially adopted for high-luminosity cases at high redshifts in recent near-IR spectroscopic studies (e.g., \citealp{Jiang07,Yi14,Wang15,Zuo15,Wu15}). Furthermore, the emission-line width of \mgii~ has been demonstrated to correlate with that of H$\beta$ in single-epoch spectra (e.g., \citealp{Salviander07, Shen08, McGill08, Wang09, Shen16}), so it has become a preferred estimator of BH masses from intermediate redshifts to the most-distant quasars (e.g., \citealp{Shen11,Willott10,Trakhtenbrot12,Zuo15}).

\begin{figure}[h]
 \includegraphics[width=8.6cm, height=5.5cm, angle=0]{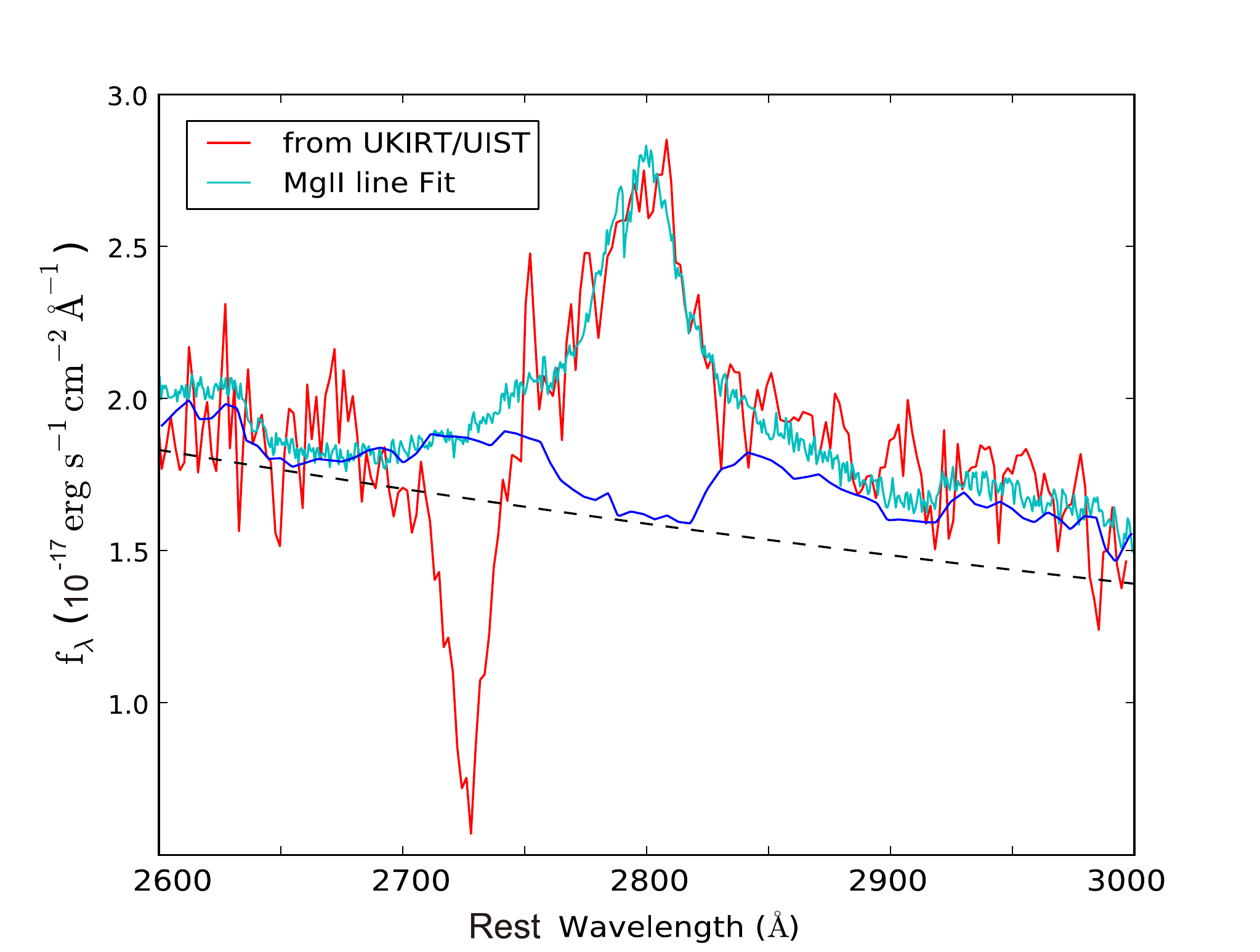}
 \caption{The cyan curve presents the spectral fit of the observed spectrum (red line), including the \mgii\ line, continuum (black dashed line) and \feii\ components (blue line).   }
  \label{fig5}
\end{figure}

The empirical single-epoch relation usually adopted to estimate SMBH masses from the Mg II line and  its associated continuum at 3000~\AA\ is as follows:\\
\begin{equation}
    M_{\rm BH,vir}=10^{a}\times (\frac{\lambda L_{\lambda}}{10^{44} {\rm erg\, s^{-1}}})^{b}\times (\frac{\rm FWHM}{\rm km\, s^{-1}})^2
\end{equation}
This relation, in fact, is derived from a virial assumption of gaseous in the broad emission line region. Therefore, both $a$ and $b$ are determined by the virial coefficient $f$. This factor $f$ could be calibrated by an independent way (such as the empirical $M-\sigma$ relation) to a higher confidence level. Therefore,  finding reliable measurements for the \mgii~ line and the $f$ factor would be critically important to estimate the BH mass. However, different fits among these broad emission lines \citep{Wang09}, different $f$ factors between classical bulges and pseudobulges \citep{Ho14} , and a potentially new $R_{BLR}-L$ relation depending upon accretion rates \citep{Du16}, should be taken into account when estimating the BH mass, which in turn, make it hard to mitigate intrinsic uncertainties of the single-epoch spectral relation. 
For simplicity, $a=0.86$ and $b=0.5$ are adopted in this paper respectively, according to the study from \citet{Vestergaard09}.

Though the \mgii\ absorption line appears on the blueward part of the emission line, the measurement of FWHM should be reliable from the symmetric profile fit of the emission line. 
Here, we calculated the 3000\AA~ monochromatic luminosity ($L_{\rm 3000\AA~}\sim5.81\times10^{46}$ \ergs) based on the analysis of the near-IR spectrum after correction for Galactic extinction \citep{Schlafly11}.
We then assume a bolometric correction factor of $\sim 5.15\times L_{\rm 3000\AA~}$ \citep{Richards06} to obtain the bolometric luminosity after subtracting the UV Fe II contribution. 
 Noticing random atmospheric changes may lead to a large uncertainty for near-IR spectral flux calibration, an alternative method based on the luminosity relation ($L_{\rm 3000\AA~} \sim 1.62L_{\rm 5100\AA~}$, see \citealp{Trakhtenbrot12}) is also taken into account. 
Since \mgii\ BALs are more frequently detected in quasars with narrower H$\beta$ and weaker \oiii\ emission lines, stronger UV and optical \feii\ multiplets, and higher luminosity (e.g., \citealp{Zhang10}), we neglect the \oiii\ emission contribution on the marginal blueward end of the W1 bandpass. This is also consistent with the anti-correlation between \feii\ and \oiii\ strength (\citealp{Boroson92}). Then $L_{\rm 3000\AA~} \sim 6.19\times10^{46}$ \ergs~ was derived. 
Using the empirical single-epoch relation with FWHM of the \mgii\ line  $\sim$3600 km s$^{-1}$ and the 3000~\AA~ monochromatic luminosity, the BH mass is estimated to be $\sim 2.3 \times 10^9 M_\odot$. 
The Eddington ratio, a non-dimensional physical parameter defined by the bolometric luminosity over its Eddington luminosity, is $\sim 1.0$ based on the flux calibration and spectral fit (see Fig. \ref{fig5}).

A BAL QSO such as this with extreme outflows might have additional velocity structure even in \mgii\  in addition to the gas motions dominated by gravitation that are assumed to provide the measure of the black hole mass \citep{Plotkin15}. 
We also note the possibility of systematic error in the \mgii\ estimator for very high luminosities that is assumed to be the same as that for the low-luminosity QSOs that have been reverberation mapped.
In such a case, the inherent BH mass uncertainty of J0122+1216 would be higher than that of the single-epoch spectral relation ($\sim 0.3$ dex) according to previous studies at low and intermediate redshifts (e.g., \citealp{Vestergaard09,Wang09,Shen11}).

\subsection{Determining the BI(\mgii)}
To calculate the BI, we first normalize the spectrum by dividing by the continuum fit to the observed spectrum in less-absorbed spectral regions around 1700 \AA~ and to the red of 3500 \AA~ in the rest frame. 
Combining the continuum derived from optical and near-IR spectra provides the framework to analyze absorption troughs quantitatively for both the \mgii\ and \civ\ lines. 

\begin{figure}[h]
 \includegraphics[width=8.6cm, height=5cm, angle=0]{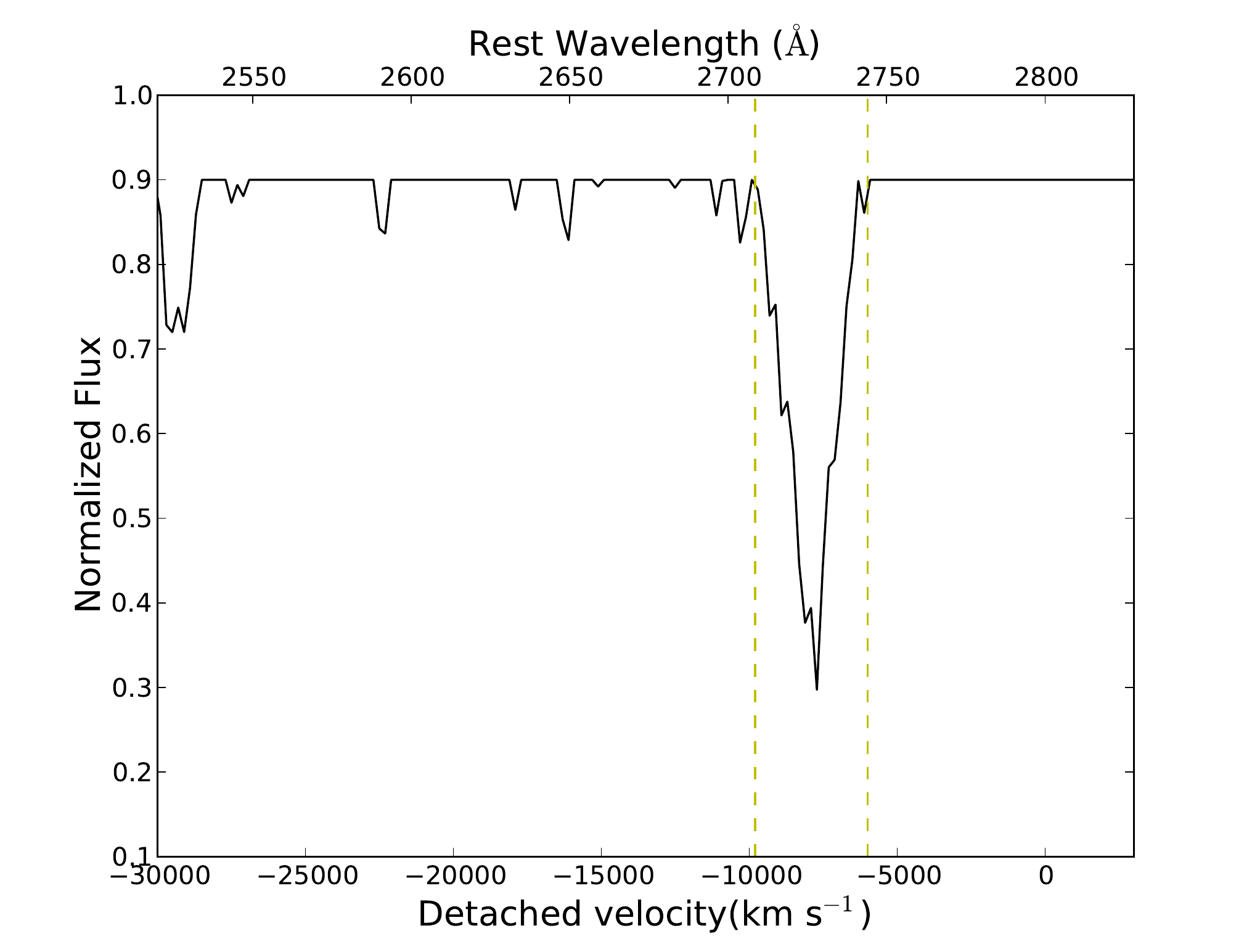}
 \caption{The normalized spectrum of this LoBAL QSO around the \mgii\ line, in which only the wavelengths with normalized flux below 0.9 times the  continuum level were used in the calculation of the BI index between the yellow lines.}
  \label{fig4}
\end{figure}
For the LoBAL QSO J0122+1216, with the detached \mgii~ absorption trough far away from its emission peak, we adopt the traditional absorption BI to be consistent with the discovery paper, which is defined as follows (\citealp{Weymann91}):
\begin{equation}
   {\rm BI} = \int^{max}_{min}[1-\frac{f(v)}{0.9}]C{\rm d}v
\end{equation}

$f(v)$ is the continuum-normalized spectral flux at a velocity $v$ (in \kms) in the rest frame. The dimensionless value $C$,  is set to 1 at velocity width more than 2000 km $\rm s^{-1}$ from the start of a continuous trough with flux density less than 90\% of the continuum, and $C = 0$ elsewhere.
In addition, we define velocities flowing outward along our line of sight to be negative, which are distinguished from the QSO emission rest frame (see Fig \ref{fig4}).
Here, the 3000 km s$^{-1}$ red limit and 10000 km s$^{-1}$ blue limit are set to avoid absorption lines unrelated to the \mgii~ ion. Further, the \mgii~ velocity components are not seen at higher velocities. 
Then, the BI(\mgii) is calculated to be $\sim$ 1350 \kms, with the main uncertainty being the determination of the intrinsic continuum.

\subsection{Broad-Band Spectral Energy Distribution}
\begin{table*}[!]
\caption{Observational properties derived from the spectra compared with other high-$z$ BAL QSOs}
\begin{tabular}{lllllllllr}
\hline
\\
Name & $z$ & type & BI(CIV) & (\civ)$^a$ & (\mgii)& (\hei~) & $\alpha_\lambda $$^b$ & SDSS-$i$ & $\rm M_{i^*}$$^d$   \\
&&&\kms&position&REW$^c$ &REW&&&\\
\\
\hline
(From \citealt{Maiolino04a})\\
SDSSJ104845.05+463718.3 & 6.22 & LoBAL & 6500$\pm$1100 & 11000/20000 &  && $-$2.10  &  22.4 & --27.4 \\
SDSSJ104433.04-012502.2 & 5.78 & HiBAL & 1950$\pm$250  & 5000 &  && $-$1.55 & 21.6 & --28.4 \\
SDSSJ075618.14+410408.6 & 5.08 & BAL & 270$\pm$60    & 9000 &  & &$-$1.67 &  20.2 & --26.6 \\
SDSSJ160501.21$-$011220.6 & 4.92 & LoBAL$^e$ & 9300$\pm$2000 &11000 &  && $-$1.35 & 19.8 & --27.7 \\
\hline
(Our Target)\\
SDSSJ012247.34+121624.0 & 4.82$^h$ &   LoBAL    &     15300$\pm$2000$^f$    &12000/21000 & 14.64 & 3.54 & $-$2.02 &  19.4 &  --27.9 \\
\\
\hline
\end{tabular}
\vspace{0.1cm}\\
Notes:\\
$^a$ The deepest position of \civ\ absorption troughs in the unit of \kms. Two of them have apparent double-trough components.\\
$^b$ Slope of the continuum ($\rm F_{\lambda}\propto \lambda ^{\alpha }$); this is calculated by comparison with the non-BAL SDSS template which has $\alpha  = -1.6$.\\
$^c$ REW is the rest equivalent width.\\
$^d$ Absolute magnitude in the rest frame i$^*$ band, obtained by using the observed spectral slope and assuming
$\rm H_0 = 50~km~s^{-1}~Mpc^{-1}$,
$\rm \Omega _m=1$ and $\rm \Omega _{\Lambda}=0$ (to allow for a direct comparison with \citealt{Reichard03a}); it is not corrected
for intrinsic luminosities of the QSOs.\\
$^e$ Possibly a FeLoBAL.\\
$^f$ All errors on the balnicity indices are dominated by the uncertainty on the profile of the CIV emission.\\
$^g$ The UV continuum of this LoBAL may be fitted with dust reddening, but with an extinction curve deviated 
 to the SMC.\\
$^h$ This is the redshift re-identified by the \mgii\ emission-line, while the redshift based on the optical spectrum reported from the discovering paper is $z=4.76$.
\label{tab3}
\end{table*}

\begin{figure*}[!]
 \includegraphics[width=19cm, height=5.5cm, angle=0]{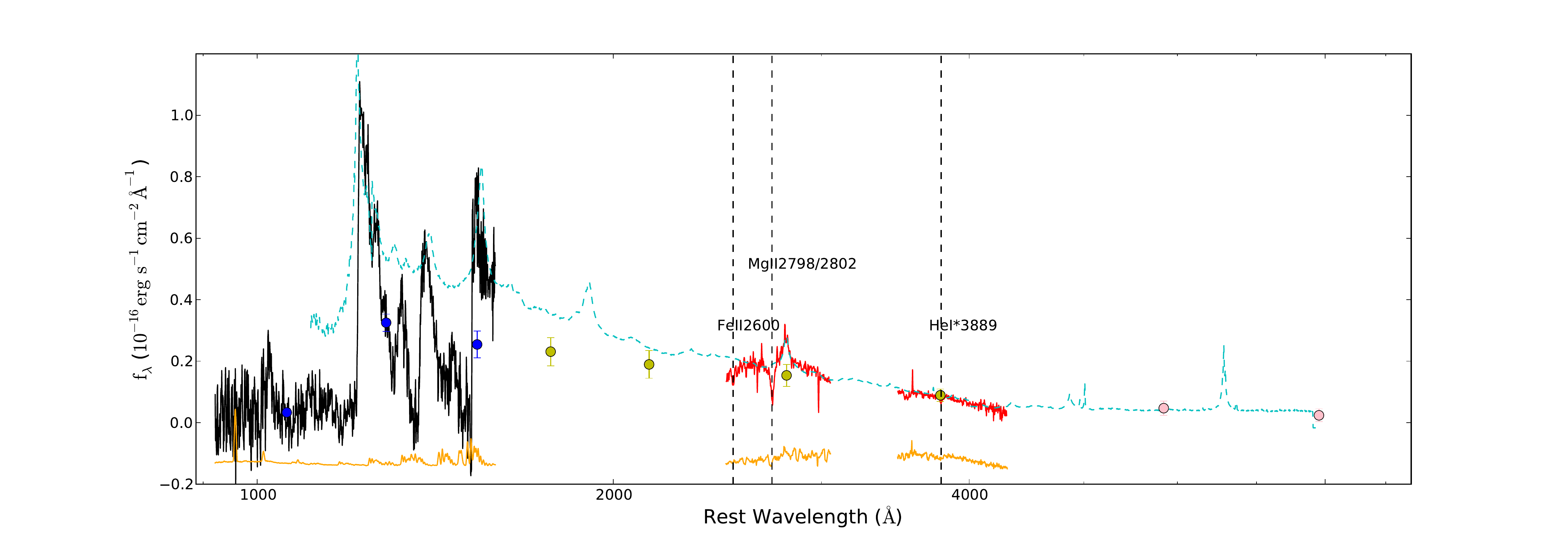}
 \caption{Applying SMC-like and the mean extinction curves at $z > 4$ (\citealp{Gallerani10}) to the intrinsic spectra with slope of $-2.02$, we get the spectral fitting (cyan dashed line) of this LoBAL. The the orange line on the bottom is the uncertainty distribution of the spectral flux. Three vertically black dashed lines mark the positions of three low-ionization absorption troughs. Blue/yellow/pink dots are the photometric data from SDSS, UKIDSS, {\it WISE} sky surveys, respectively. }
  \label{fig7}
\end{figure*}

Extensive studies of interstellar attenuation from reddening and scattering have been done, but most of them are focused on the local universe. 
Generally, the SMC extinction curve is steeper than those of the LMC, the MW, and starburst galaxies \citep{Calzetti94}. However, 
some individual studies showed that extinction curves for LoBAL QSOs may be steeper than the SMC extinction curve shortward of 2000 \AA~ (e.g., \citealp{Hall02}). In this scenario, a smaller average grain size is needed to produce steeper extinction curves, since small particles cannot scatter efficiently at long wavelengths. In contrast, the flat extinction curves in the circumnuclear regions of Seyfert galaxies could be interpreted as being produced from dust dominated by very large grains.

To match the observed spectra, different quasar templates were adopted mainly by adjusting the continuum slope ($\alpha_\lambda$), the absolute extinction ($A_{3000}$), and by applying some empirical and theoretical extinction curves.
The average intrinsic slope ($\alpha_\lambda$) of BAL QSOs is about $-2.02$ at low redshifts \citep{Reichard03a}, so we took an unreddened QSO template with this slope to fit the continuum sections at $\lambda > 3000$ \AA, which would be less affected by intrinsic extinction. 
In order to match the composite quasar spectrum to the input spectra, we allow for changes in the slope and shape of the continuum between the input quasar spectrum and the template quasar spectrum, and the spectrum of the LoBAL is fitted with the following expression:
\begin{equation}
F_{\lambda}=C~F_{\lambda}^{t}~\left ( \frac{\lambda}{\rm 3000\AA}\right )^{(1.62+\alpha_{\lambda})}~10^{-\frac{A_{3000}}{2.5}
\frac{A_{\lambda}}{A_{3000}}},
\label{bfeq}
\end{equation}
where $C$ is a normalization constant, $F_{\lambda}^{t}$ is a quasar template spectrum, $\alpha_{\lambda}$ is the intrinsic slope
of the unreddened spectrum, and $A_{\lambda}/A_{3000}$ is the extinction curve
normalized at 3000~\AA, which is derived from empirical and theoretical extinction curves (e.g., \citealp{Maiolino04a,Gallerani10}). 
The extinction at a wavelength $\lambda$ is related to the color excess $E(B-V)$ and to the reddening curve by:
\begin{equation}
A_{\lambda} = k_{\lambda}E(B-V) =  \frac{k_\lambda A_V}{R_V}
\label{bfeq}
\end{equation}

Here, we adopt the template compiled by \citet{Reichard03a}, obtained using 892 quasars classified as non-BAL, whose continuum is described by a power-law with average index $\alpha_{\lambda,t}\sim-1.6$
($F_{\lambda}^{t}\propto \lambda^{\alpha_{\lambda,t}}$). We change the slope of the template with the term $(\lambda/3000)^{1.62+\alpha_\lambda}$. 
The redward part of the spectra at $\lambda_{rest} >3000$ \AA~ is the least-extinguished region and so is used to anchor the reddening fit. 
However, we did not include $2500<\lambda<3000$ \AA~ in the fit since it has the prominent \feii\ bump and absorptions as shown in Figure \ref{fig5}. 
Meanwhile, we exclude the spectral format edges ($<2500$ \AA~ and $>4300$ \AA~) from the reddening fit.
The continuum emission peak between \siiv\ and \lya\ was also excluded because it falls into the strong atmospheric absorption region around $\sim$7600 \AA~.   
Adopting an intrinsic spectrum with slope of $-2.02$ and an absolute extinction at 3000 \AA~ of 0.35 mag, the blue-part continuum slope of J0122+1216 is estimated to be $\sim -1.6$, which is very close to that of the average composite spectral slope. 
In principle, the slope of the optical spectrum, from the \mgii\ emission-line profile redward in the observed-frame near-IR spectrum to rest wavelengths larger than 3500 \AA~,  could be used to visually check the consistency of our fit to the whole observed spectra (see the cyan dashed line in Fig. \ref{fig7}). 
However, for a given extinction curve $f_\lambda = A_\lambda / A_{3000}$, any other extinction curve of the form $f_\lambda '= k( f_\lambda - 1) + 1$, can fit the observed spectrum equally well by simply changing the absolute extinction $A_{3000}$. Thus, we cannot eliminate a degeneracy in the slope of the extinction curve without knowledge of the intrinsic luminosity.

On the other hand, adopting a lower uncertainty factor of $\sim$2 from the single-epoch spectral relation, the maximum BH mass and intrinsic bolometric luminosity are $\sim4.6\times10^9 M_\odot$ and $\sim 6.3\times10^{47}$ \ergs~ constrained by the conventional Eddington limit, which corresponds to a monochromatic luminosity of $1.2\times10^{47}$ \ergs~ at 3000 \AA~. According to these values, the maximum absolute extinction $A_{3000}$ should not be larger than $\sim 0.72 $ mag giving an intrinsic continuum slope not steeper than $-2.8$. Very few QSOs have been found with such a blue UV slope,
and a new Eddington limit ($4.4\times10^{47}$ \ergs) could be derived after one iteration using the monochromatic 3000 \AA~ luminosity and relation (1). 
Thus we adopt the values of intrinsic continuum slope of $-2.02$ and 0.35 mag extinction at 3000 \AA~ as reasonable and matching the observed spectra, application of which yields the final spectral fit shown in Figure \ref{fig7}. 

The assumptions discussed above were made for a reasonable case since it is hard to determine the exact reddening when considering the degeneracy between $\alpha_\lambda$ and $A_V$. Among the repeated fits of the intrinsic and observed spectra, the SMC extinction curve seems to deviate substantially from that of the LoBAL QSO J0122+1216 when considering the fact that it will make the whole spectrum redder. 
However, a detailed investigation of the extinction properties of J0122+1216 would require higher S/N spectral data obtained in simultaneous observations  and wider wavelength coverages, which is beyond the scope of this paper.

\subsection{Physical constraints on the outflows}

AGN feedback seems to play an important role in controlling the co-evolution between central SMBHs and host galaxies over a wide range of cosmological scales, and thus it is essential to understand the detailed structure and physics of outflows (e.g., \citealp{Di05}). Generally, both emission and absorption lines can be efficient diagnostics of the geometry and dynamics of quasars. 
To be specific, the emission lines 
are usually used to determine the black-hole mass or even to derive a global covering factor of outflows;  the absorption lines, which are associated with  outflows in the line of sight, are good for determining the total hydrogen column density ($N_H$) and the velocity distribution of the outflows \citep{LiuW15}.

\subsubsection{The high-ionization \civ\ emission-line}

Since the velocity field and gas density distribution cannot be determined according to the analyses of the optical and near-IR spectra available, an alternative way to estimate the kinetic power of J0122+1216 is necessary.
The dramatic blueshifts of high-ionization lines as shown in Figure \ref{fig2} provide evidence that the \civ\ line region may be dominated by outflows rather than the gravitational virial motions prevalent in most normal quasars at low redshifts \citep{Wang11}, or at least, these blueshifted lines are partially caused by outflows. 
Therefore, a simplified method based on an outflow model was adopted to obtain an order of magnitude estimate of the kinetic power associated with the \civ\ emitting outflows under several assumptions. 
We start by considering the fraction of the flux in the  \civ\ blue component that is above the expected projected escape velocity at $r \sim 1500$ $r_g$. This radius   ($\sim10$ times larger than that of the \civ\ broad emission line region) is set to make sure it will cover the relevant broad emission line regions.

The outflowing kinetic power, at the velocity of $v$\ in units of 5000 km s$^{-1}$ and the integrated \civ\ line luminosity in units of \ergs\  \citep{Marziani16}, is: 
\begin{equation}
 \dot{\epsilon}(v) = \frac{1}{2} \dot{M}^\mathrm{ion}_\mathrm{out} v^{2} \approx 1.2 \cdot 10^{44} L_{45} v^{3}_{5000}   r^{-1}_{\rm 1} \rm  (erg~s^{-1})
\end{equation}
where $r^{-1}_{\rm 1}$ is the inner radius of the \civ\ emission-line outflowing region (here $r_1 = 1500r_g \sim 2.8$ pc is assumed).   
First, we measure the integrated \civ~ line luminosity ($\sim4.3\times10^{45}$ \ergs) after subtraction of the continuum. 
From the blueshifted side of the \civ~ line ($\sim$5000 \kms~) which is likely driven by outflows, the kinetic power associated with the \civ~ line can be roughly estimated to be $\sim$4.0 $\times 10^{45}$ \ergs~ ($\sim$1.2\% L$_{Edd}$). 
We note that inner radius of $1500r_g$ and an electron density of $10^9$ cm$^{-3}$ are assumed to derive equation (5), but both of them are likely linked to lower limits when compared with physical properties of the \civ~ emission-line region.  Thus, adopting this simplified approach, the outflow kinetic power may be underestimated to some extent.

\subsubsection{Analysis based on absorption lines}
As a comparison, we also follow a conventional procedure to analyze the \civ\ , \hei3889 and \mgii\ absorption troughs, which is helpful to study outflowing material along our line of sight. 
For an ion giving rise to an unsaturated absorption line, the optical depth at velocity $v$ is directly proportional to the column density corresponding to $v$ (here we use $N_v$ to represent this physical quantity). Then, the total column density $N$, in units of cm$^{-2}$, can be approximated by  
\begin{equation}                 \label{e_N}
N = \int N_v dv = \sum N_v \Delta v = {{3.7679 \times 10^{14} }\over{\lambda f_{jk}}} \sum \tau_v \Delta v
\end{equation}              
where $f_{jk}$ is the oscillator strength corresponding to a specific ion, $\tau_v$ is the optical depth at velocity $v$, and $\Delta v$ is the velocity width element in the center of $v$. 
Using Equation \ref{e_N}, we derived the \hei~ column density to be $4.5\times 10^{14}$ cm$^{-2}$, adopting the oscillator strength of 0.064, a velocity width of 2000 \kms~ and an apparent optical depth of 0.15.

The lower limit on the total hydrogen column density ($N_H$), which tightly depends on the ionization states and abundances, can also be estimated to some extent. Because there are still substantial uncertainties that depend on results from photoionization simulations, we perform only a qualitative analysis here. 
First,  the \hei~ absorption line multiplets can be used as probes to obtain the real ion column density as opposed to a lower limit available from the apparent column density \citep{Arav01}. Then, the column density of helium in the metastable state ($4.5\times 10^{14}$ cm$^{-2}$) derived from the \hei($\lambda3889$) absorption line can yield a conservative lower limit on the ionized helium column density and hence the ionized hydrogen column density ($N_{H II} \sim 10^{21}$ cm$^{-2}$ ). 
Furthermore, a lower limit on the hydrogen column density ($N_H = N_{H I} + N_{H II}  > 10^{22.5}$ cm$^{-2}$) can be obtained according to the detailed photoionization simulations for reasonable parameter ranges (e.g. \citealp{Leighly11}). The column density estimated above, in fact, is also corroborated by other observational features such as a LoBAL with \feii~ absorption that would require high column densities (e.g., \citealp{Mudd16}).

With a lower limit on the total hydrogen column density ($3.2\times 10^{22}$ cm$^{-2}$), the fraction of outflowing kinetic power can be somewhat constrained. 
The relation of outflowing kinetic power is expressed as follows:
\begin{equation} \label{k_E}
  \dot{\epsilon} = \frac{1}{2}\dot{M}^\mathrm{ion}_\mathrm{out} v^{2} = 2\pi \mu m_p R\Omega N_Hv^3
\end{equation}
where $\mu m_p$=1.4$m_p$ is the mean mass per proton in the absorber, and $\Omega$ is the global covering factor.

For the reliably detected \mgii\ BALs of J0122+1216 at $z=4.82$, a mean absorbed flux-weighted centroid velocity of 8000 \kms , an assumed value of $R=60$ pc (e.g., \citealp{Ji15,LiuW15}), and a global covering factor of $\Omega=0.1$ were adopted according to investigations of LoBAL QSOs at $z<1$ (e.g., \citealp{Trump06, Zhang10, LiuW15}). The lower limit on kinetic power is estimated to be 2.8 $\times 10^{45}$ \ergs~ ($\sim$0.9\% of L$_{Edd}$) using Equation (\ref{k_E}). Here, we adopted a lower value of $\Omega$, which is mainly based on optical spectroscopic statistics, and thus could be underestimated due to obscuration effects.

\section{Discussion}

Understanding the origin of outflows and feedbacks are fundamental to our understanding of various observational phenomena of BALs. 
Some authors have proposed that the different line types represent different manifestations of a single outflow phenomenon viewed at different angles (e.g, \citealp{Elvis00}). 
However, BAL QSOs only appear in a small fraction of the observed spectra, indicating that either they are only seen in particular directions with respect to the axis of the accretion disk or they are observed during particular phases of the BAL QSO life (Farrah et al. 2007, and references therein). Thus, the possibility still remains that a combination of evolutionary and orientation effects can explain the separation of BAL QSOs and non-BAL QSOs. For example, the structure of the outflows could change either with cosmic time or an orientation dependence, but it is difficult or impossible to favor one of them based on observations of individual objects.

A redshift dependence of the broad absorption line quasar fraction has been found by Allen et al.(2011), implying that an orientation effect alone is not sufficient to explain the presence of BAL troughs in some but not all quasar spectra. 
The outflows giving rise to these broad absorption lines once had been thought to originate in accretion disk instabilities and are accelerated to velocities of up to $0.2c$ in the quasar rest-frame \citep{Hamann13} by magnetic effects (e.g., \citealp{Everett05}) or thermal winds (e.g., \citealp{Giustini12}) from the accretion disk. For the LoBAL J0122+1216, the idea of powerful radiation pressure driven by the central SMBH seems to be plausible in terms of the outflows appearing among a fairly large fraction of luminous QSOs (e.g., \citealp{Arav95,Proga00,Chelouche01}), especially noticing its high luminosity and Eddington ratio.

\citet{Boroson02} found that BAL QSOs on average have higher Eddington ratios and accretion rates than those of non-BAL QSOs in a small sample of BAL QSOs, and a similar conclusion has been corroborated by recent studies (e.g., \citealp{Ganguly07, Zubovas13}). 
Moreover, some studies suggest that BAL QSOs are redder and/or more luminous than other quasars \citep{Reichard03a, Trump06, Gibson09}. 
The high Eddington ratio of the LoBAL QSO J0122+1216, in fact, fits into this scenario of violently accreting systems having stronger outflows.
Further, observational evidence to support the luminosity - outflow connection has been found between the blueshift and asymmetry of the \civ\ profile and the Eddington ratio (e.g.,\citealp{Laor02,Wang11}). 
It is possible that high Eddington ratios, even though not the main driver in the feedback processes of BAL QSOs, may control the formation of extreme outflows with sub-relativistic components under extreme circumnuclear conditions. 
 Though the variability of Ly$\alpha$ and Ly$\beta$ emission lines has been detected, no corresponding pattern with the variability of \civ\ or \siiv\ absorption troughs has been found. Thus, due to the effects of low S/N spectra and a short time span of monitoring, we cannot establish a relation between the high-ionization broad emission/absorption lines to pin down the origin of circumnuclear outflows. 

According to theoretical models, at least 0.5\% of $L_{\rm Edd}$ is required to provide substantial feedback to the host galaxy in the form of outflowing material (e.g., \citealp{Di05, Hopkins05, Hopkins10}). The lower limit on kinetic power estimated from both emission and absorption lines of J0122+1216 are already larger than that required for significant AGN feedback. 
Due to the uncertainties of estimating the kinetic power of outflows, which are typically one order of magnitude or even more, we caution that the method adopted in this paper may be more uncertain given all the assumptions adopted. However, a lower limit on the outflowing kinetic power of J0122+1216 can be safely set through the analyses mentioned above. 
Thus, we find supporting observational evidence that the circumnuclear outflows of J0122+1216 could contribute significantly to the origin of galactic-scale feedback effects, based on the measurements of the BH mass and bolometric luminosity. 

\section{Conclusion}

The quasi-ubiquitous outflows in quasars may play an important role on sub-parsec as well as kilo-parsec scales in controlling the growth of the central black hole, the evolution of the host galaxy, and the chemical enrichment of the intergalactic medium according to observational and theoretical predictions (e.g., \citealp{Moe09,Ostriker10}), yet more observational evidence is needed, especially for high-redshift cases.

There are various methods being adopted to learn about the evolution of quasars, however, studying absorption lines from the outflows of BAL QSOs  is the most direct and efficient way to learn their geometrical structures and chemical abundances, kinematics and extinction, or even more fundamental physics in terms of the accretion disk, the powering mechanism, the feedback and growth of SMBH processes etc. in the early Universe.
In this paper, the follow-up investigations of this newly discovered LoBAL QSO have been made from an observational view, and we briefly conclude our  findings and suggested implications as follows:

1)Based on near-IR spectra with  \mgii\ absorption/emission lines, J0122+1216 was identified to be a LoBAL QSO, and its redshift was determined to be 4.82 instead of 4.76. With this systemic redshift, the high-ionization UV emission lines present large blueshifts, which could be directly associated with extreme activity of circumnuclear outflows. 

2)During 16 months of monitoring, the variability of \civ\ absorption troughs is small after spectrophotometric calibration, while it is conspicuous for the Ly$\alpha$ and Ly$\beta$ emission lines, which again, could be an observational signature characterizing extreme outflows.

3)The \mgii-based BH mass of this LoBAL QSO is about 2.3 $\times10^9$ M$_\odot$ according to the empirical single-epoch spectral relation. However, the intrinsic deviation from the relation could be larger than the typical 0.3 dex, in terms of these dramatic outflows launching from the high-luminosity region at such a redshift.

4)We find observational evidence that the Eddington ratio of this LoBAL QSO is very close to or even beyond the conventional limit, which may become the main driver of its sub-relativistic outflows under extreme circumnuclear environments.

Outflows are generally believed to be fundamental to understand the process of AGN feedback, so a better understanding of their geometries and kinematics could shed light on the physical basis of BAL QSOs. 
Considering the fact that large variabilities of BALs usually happen during a relatively long time and the typical life time of outflows is longer than several tens of years in the rest-frame for most BAL QSOs, it is meaningful to put this newly discovered LoBAL QSO in a long-term monitoring campaign for further study. In addition, some observational findings, (e.g., X-ray emission from BAL AGNs is, on average, much weaker than non-BAL AGNs, and their circumnuclear regions tend to be more compact and polarized), are of great interest to advance our understanding of BAL QSOs. 
Therefore, more extensive follow-up observations would be helpful to confirm these findings and further reveal intrinsic physics of the LoBAL QSO J0122+1216.

\acknowledgments

We acknowledge stimulating discussions with Shaohua Zhang, Wenjuan Liu and Xiaobo Dong. We acknowledge the support of the staff of the Lijiang 2.4m telescope (LJT). Funding for the telescope has been provided by CAS and the People's Government of Yunnan Province. W. Yi thanks the support from the West Light Foundation of The Chinese Academy of Sciences (Y6XB016001). W. Yi also thanks the financial support from the program of China Scholarships Council No. 201604910001 for his postdoctoral study at the Pennsylvania State University. CJG and WNB acknowledge support from NSF grants AST-1516784 and AST-1517113. 
The United Kingdom Infrared Telescope (UKIRT) is supported by NASA and operated under an agreement among the University of Hawaii, the University of Arizona, and Lockheed Martin Advanced Technology Center; operations are enabled through the cooperation of the East Asian Observatory. This research uses data obtained through the Telescope Access Program (TAP), which has been funded by the National Astronomical Observatories of China, the Chinese Academy of Sciences (the Strategic Priority Research Program "The Emergence of Cosmological Structures" Grant No. XDB09000000), and the Special Fund for Astronomy from the Ministry of Finance. Observations obtained with the Hale Telescope at Palomar Observatory were obtained as part of an agreement between the National Astronomical Observatories, Chinese Academy of Sciences, and the California Institute of Technology.
We used the quasar spectral fitting code kindly provided by Yue Shen. 
We also acknowledge the assistance from Antonio Magazzu, Emilio Molinari and Thomas Augusteijn at the TNG and NOT observatories.


{\it Facilities:} \facility{The Lijiang 2.4m Telescope (LJT)}, \facility{YFOSC}, \facility{TNG}, \facility{NICS}, \facility{UKIRT}, \facility{UIST}, \facility{P200}, \facility{TripleSpec}.

\end{document}